\documentclass[aps,prd,preprint,superscriptaddress]{revtex4-2}

\usepackage{ulem}
\usepackage{xcolor}
\usepackage[colorlinks=true]{hyperref}
\usepackage{soul}
\usepackage{booktabs}
\usepackage{mathrsfs}
\usepackage{amssymb}

\usepackage{amsmath,amssymb,color,epsfig}
\allowdisplaybreaks[4]

\newcommand{\be}{\begin{equation}}
\newcommand{\ee}{\end{equation}}
\newcommand{\bea}{\setlength\arraycolsep{2pt} \begin{eqnarray}}
\newcommand{\eea}{\end{eqnarray}}
\newcommand{\nn}{\nonumber}

\def\0{{\sst{(0)}}}
\def\1{{\sst{(1)}}}
\def\2{{\sst{(2)}}}
\def\3{{\sst{(3)}}}
\def\4{{\sst{(4)}}}
\def\5{{\sst{(5)}}}
\def\6{{\sst{(6)}}}
\def\7{{\sst{(7)}}}
\def\8{{\sst{(8)}}}
\def\sst#1{{\scriptscriptstyle #1}}

\begin{document}

\hypersetup{
    linkcolor=blue,
    citecolor=red,
    urlcolor=magenta
}


\title{Dyonic RN-like and Taub-NUT-like black holes in Einstein-bumblebee gravity}




\author{Shoulong Li}
\email[]{shoulongli@hunnu.edu.cn}
\affiliation{Department of Physics, Key Laboratory of Low Dimensional Quantum Structures and Quantum Control of Ministry of Education, and Institute of Interdisciplinary Studies, Hunan Normal University, Changsha, 410081, China}
\affiliation{Hunan Research Center of the Basic Discipline for Quantum Effects and Quantum Technologies, Hunan Normal University, Changsha 410081, China}

\author{Liang Liang}
\affiliation{Department of Physics, Key Laboratory of Low Dimensional Quantum Structures and Quantum Control of Ministry of Education, and Institute of Interdisciplinary Studies, Hunan Normal University, Changsha, 410081, China}
\affiliation{Hunan Research Center of the Basic Discipline for Quantum Effects and Quantum Technologies, Hunan Normal University, Changsha 410081, China}

\author{Liang Ma}
\email[Corresponding author: ]{maliang0@tju.edu.cn}
\affiliation{Center for Joint Quantum Studies and Department of Physics,
School of Science, Tianjin University, Tianjin 300350, China}


\date{\today}

\begin{abstract}

Einstein-bumblebee gravity is one of the simplest vector-tensor theories that realizes spontaneous Lorentz symmetry breaking. In this work, we first construct an exact dyonic Reissner-Nordstr\"om-like black hole solution in four dimensions, carrying both electric and magnetic charges and admitting general topological horizons. We then study its thermodynamic properties, and employ the Wald formalism to compute the conserved mass and entropy, thereby establishing the first law of black hole thermodynamics. Furthermore, we generalize these results to Taub-Newman-Unti-Tamburino case and  higher dimensions case.
 
\end{abstract}


\maketitle

\section{Introduction}

Lorentz symmetry, or Lorentz invariance, is a fundamental postulate of both Standard Model and general relativity (GR). Over the past decades, its validity has been extensively investigated from both theoretical and experimental perspectives~\cite{Mattingly:2005re,Liberati:2013xla,Will:2014kxa,Berti:2015itd,Hees:2016lyw,Wei:2021vvn,Li:2025yvq}. From the theoretical side, it has been suggested that Lorentz invariance might not be an exact symmetry at all energy scales. In this context, a variety of quantum gravity models accommodating Lorentz violation have been proposed, including string theory~\cite{Kostelecky:1988zi,Ellis:1999yd}, warped brane worlds~\cite{Burgess:2002tb}, and loop quantum gravity~\cite{Gambini:1998it}. If Lorentz invariance is broken, one naturally expects significant violation at the Planck scale, around $10^{19}$ GeV, while small residual effects could appear at low energies. Furthermore, even if a quantum gravity theory preserves Lorentz invariance at the Planck scale, it may contain tensor fields that acquire nonzero vacuum expectation values (VEVs) at lower energies, thereby spontaneously breaking the symmetry~\cite{Kostelecky:1988zi}. With steady advances in experimental techniques, the search for low-energy signatures of Lorentz violation remains an active and enduring line of research.

Based on the idea of spontaneous Lorentz symmetry breaking in string theory~\cite{Kostelecky:1988zi}, Colladay and Kosteleck\'y~\cite{Colladay:1996iz,Colladay:1998fq,Kostelecky:2003fs} proposed a model-independent framework, known as the Standard Model extension, to systematically incorporate almost all possible Lorentz-violating effects in both the Standard Model and GR. In the gravitational sector, many important modified gravity theories, particularly those involving coupled vector fields, can be formulated within the framework of Standard Model extension. For a comprehensive review, we refer the reader to Ref.~\cite{Liberati:2013xla} and references therein. 

Among these Lorentz-violating gravity theories, Einstein-bumblebee gravity~\cite{Kostelecky:2003fs,Bailey:2006fd} represents one of the simplest vector-tensor models that realizes spontaneous Lorentz symmetry breaking. Since its proposal, it has attracted sustained research interest. In particular, Casana et al.~\cite{Casana:2017jkc} constructed an elegant Schwarzschild-like exact black hole solution, which allows one to conveniently study the implications of spontaneous Lorentz symmetry breaking in the gravitational sector through black hole physics and astrophysical phenomena, thereby further stimulating attention to the theory. Subsequently, several exact black hole solutions, analogous to their GR counterparts, have been constructed within Einstein-bumblebee gravity, including Schwarzschild-like black hole with a cosmological constant~\cite{Maluf:2020kgf}, electrically charged Reissner-Nordstr\"om (RN)-like black holes~\cite{Liu:2024axg}, Taub-Newman-Unti-Tamburino (Taub-NUT)-like black holes~\cite{Chen:2025ypx}, high dimensional Schwarzschild-like black holes with a cosmological constant~\cite{Ding:2022qcy}, as well as a variety of other analytic and numerical solutions~\cite{Ding:2019mal,Ding:2023niy,Xu:2022frb,Santos:2014nxm,Jha:2020pvk,Gullu:2020qzu}. The properties of these black holes have also been investigated~\cite{Liang:2022gdk,Xu:2023xqh,Mai:2023ggs,Mai:2024lgk,Liu:2022dcn,Liu:2024oeq,Deng:2025uvp,Oliveira:2018oha,Liu:2019mls,Li:2020dln,Chen:2020qyp,Ding:2020kfr,Kanzi:2021cbg,DCarvalho:2021zpf,Jiang:2021whw,Wang:2021gtd,Gu:2022grg,Uniyal:2022xnq,Zhang:2023wwk,Lin:2023foj,Lessa:2023yvw,Lambiase:2023zeo,An:2024fzf,Zhang:2025acq,Gao:2024ejs,Chen:2023cjd,Capozziello:2023rfv,Ge:2025xuy,Quan:2025tgz,Yu:2025brr,Ding:2025cno,Xu:2025jvk,Media:2025udn,Singh:2025hor,Singh:2025tvk,Li:2025mid}. Moreover, Einstein-bumblebee gravity has been studied in a variety of other contexts, including compact stars, cosmology, gravitational waves, and generalized formulations, and so on~\cite{Ovgun:2018xys,Ji:2024aeg,Panotopoulos:2024jtn,Neves:2024ggn, Seifert:2009gi,Liang:2022hxd,Xu:2025bvx,Maluf:2021lwh,Neves:2022qyb,vandeBruck:2025aaa,Lai:2025nyo,Jesus:2020lsv, Filho:2022yrk,Nascimento:2023auz,Delhom:2019wcm,Delhom:2020gfv,Delhom:2022xfo,Lehum:2024ovo,AraujoFilho:2024ykw,Lehum:2024ovo,AraujoFilho:2025hkm, KumarJha:2020ivj,Marques:2023suh,Bailey:2025oun,Li:2025bzo}. For a more comprehensive overview of developments in this gravity model, we refer the reader to Ref.~\cite{Kostelecky:2003fs,Bailey:2006fd} and references therein. 

In this work, we construct exact solutions for a class of dyonic RN-like black holes in four dimensions, carrying both electric and magnetic charges and admitting general topological horizons, and study their thermodynamic properties within Einstein-bumblebee gravity. We further extend the study to include the Taub-NUT case and to higher-dimensional spacetimes. The motivations are as follows. First, both electrically and magnetically charged black holes have long been of strong theoretical and astrophysical interest~\cite{Maldacena:2020skw,Chamblin:1999tk,Kubiznak:2012wp,Hartnoll:2007ih,Blandford:1977ds,Zhang:2016rli,Bozzola:2020mjx}, constructing exact dyonic RN-like solutions in Einstein-bumblebee gravity provides a convenient setting for exploring the implications of spontaneous Lorentz symmetry breaking. Second, the thermodynamics of neutral black holes in this theory exhibits notable subtleties: the usual definitions of mass (Komar and Arnowitt-Deser-Misner mass) and entropy (Wald entropy) fail to yield a consistent first law, as reported in the literatures, such as~\cite{An:2024fzf,Chen:2025ypx}. Consistency is restored only by employing more general methods, such as the Wald formalism~\cite{Wald:1993nt,Iyer:1994ys}, to properly define the conserved mass charge and entropy. Extending such an analysis to cases with Maxwell fields, topological horizons, and higher dimensions is therefore both necessary and well motivated. Third, although previous works have reported four-dimensional electrically charged RN-like black hole solutions with spherical horizons~\cite{Liu:2024axg}, which have been widely applied in black hole physics and astrophysics~\cite{Liu:2024axg,Xu:2025jvk,Media:2025udn,Singh:2025hor,Singh:2025tvk}, the previous Einstein-bumblebee-Maxwell (EbM) theory does not admit a purely magnetic black hole, while the magnetic charge of the dyonic black holes is not an independent integration constant, but fixed by the coupling constant of the theory.  In contrast, we employ an extended framework within the EbM theory family~\cite{Kostelecky:2003fs,Bailey:2006fd,Lehum:2024ovo} that admits not only purely electric or magnetic black holes, but a well-defined exact dyonic black hole solution, where the charges are true integration constants, independent of the coupling constants of the theory. This provides a foundation for further investigation of the physical properties and potential observational signatures of dyonic black holes, as well as for exploring more complex black hole solutions within a consistent Einstein-bumblebee gravity framework. In particular, the construction of exact dyonic Taub-NUT-like black hole solutions and higher-dimensional dyonic RN-like solutions within this setup further illustrates the internal consistency and robustness of the specific EbM model.

This paper is organized as follows: In Sec.~\ref{framework}, we briefly review the EbM theory family and its general equations of motion (EOMs). In Sec.~\ref{solutiond4}, we derive the four-dimensional dyonic RN-like topological black holes in a particular theory. In Sec.~\ref{Thermodynamics}, we study the thermodynamics of these black holes and employ the Wald formalism to compute the conserved mass and entropy, thereby establishing the first law of black hole thermodynamics. Furthermore, in Sec.~\ref{taub nut dyonic}, we construct the dyonic Taub-NUT-like solution and examine its thermodynamic properties using the Wald formalism.
 In Sec.~\ref{highdim} we generalize previous RN-like results to higher dimensions. Finally, We presents a summary and discussion of the work in Sec.~\ref{conclusion}.

\section{The theory and its EOMs} \label{framework}

In this section, we briefly review the EbM theory. 
The total action $\cal S$ in $D$-dimensions ($D\ge 4$) can be expressed as
\be
{\cal S} = \frac{1}{2\kappa} \int d^D x \sqrt{-g} \left( L_1 + L_2 \right) \,,  \label{EBM}
\ee
with $\kappa = 8 \pi G/c^4$, where $G$ and $c$ denote the gravitational constant and velocity of light. The Lagrangian densities $L_1$ for Einstein-bumblebee gravity sector is given by~\cite{Kostelecky:2003fs,Bailey:2006fd}
\be
L_1 = R + \gamma R_{\mu\nu}B^\mu B^\nu  -2\kappa \left(\frac14 B_{\mu\nu}B^{\mu\nu} +V(B_\mu B^\mu )\right) \,,  \label{gravity}
\ee
where $g$ denotes the determinant of the metric $g_{\mu\nu}$, $R$ the Ricci scalar,  and $R_{\mu\nu}$ the Ricci tensor. 
As a class of vector-tensor model, Einstein–bumblebee gravity introduces a vector field $B_\mu$, commonly referred to as the bumblebee field, which triggers spontaneous Lorentz symmetry breaking. Its field strength $B_{\mu\nu}$ is defined as
\be
B_{\mu\nu} = \partial_\mu B_\nu -\partial_\nu B_\mu \,. 
\ee
The parameter $\gamma$ denotes the coupling constant associated with the non-minimal interaction between the bumblebee field and the Ricci tensor. 
The potential $V$ should be minimized to obtain a stable vacuum of spacetime when the bumblebee field acquires a nonzero VEV $b_\mu$, i.e., $\langle B_\mu\rangle = b_\mu$. This nonzero VEV introduces a preferred direction in spacetime, thereby leading to the spontaneous breaking of Lorentz symmetry. Consequently, the potential can generally be expressed as
\be
V \equiv V (B_\mu B^\mu \pm b^2) \,,
\ee
where $b^2$ represents a real positive constant.
Following discussions of symmetry-breaking potentials in the literature~\cite{Kostelecky:2003fs,Bailey:2006fd,Casana:2017jkc}, the minimum of $V$ is typically chosen to vanish, which implies
\bea
V (B_\mu B^\mu \pm b^2) \big|_{B_\mu = b_\mu} &=& 0 \,, \label{potential1}\\ 
V' (B_\mu B^\mu \pm b^2) \big|_{B_\mu = b_\mu} &=& 0 \,, \label{potential2}
\eea
where $V' (x) \equiv dV(x)/dx$ and the $b_\mu$ is determined by 
\be
b_\mu b^\mu \pm b^2 = 0 \,.  \label{potential3}
\ee
The $ \pm$ signs in Eq.~(\ref{potential3}) determine whether the field $b_\mu$ is timelike or spacelike.
The electromagnetic sector $L_2$ is governed by Maxwell theory, with the Maxwell field also exhibiting a non-minimal coupling to the bumblebee field~\cite{Liu:2024axg,Seifert:2009gi,Lehum:2024ovo,Carroll:2008pk,Gomes:2009ch}, which is given by
\be
L_2 = -\frac14 F_{\mu\nu}F^{\mu\nu} +\gamma_1 B_\lambda B^\lambda F_{\mu\nu}F^{\mu\nu}  +\gamma_2 B^\mu F_{\mu\nu} B_\lambda F^{\lambda \nu}  \,,\label{action 2}
\ee
where the Maxwell field is denoted by $A_\mu$, with its field strength defined as
\be
F_{\mu\nu} = \partial_\mu A_\nu -\partial_\nu A_\mu\,.
\ee
Here, $\gamma_1$ and $\gamma_2$ are coupling constants characterizing the non-minimal interactions between the Maxwell field and the bumblebee field.
The EOMs can be derived by varying the action with respect to the gravitational field $g_{\mu\nu}$, the bumblebee field $B_\nu$, and the Maxwell field $A_\nu$. The resulting EOMs, denoted by $E_{\mu\nu}$, $E_{B}^\nu$ and $E_{A}^\nu$, are given by
\bea
E_{\mu\nu} &\equiv& G_{\mu\nu} + \gamma \bigg[-\frac12 g_{\mu\nu} R_{\rho\sigma} B^\rho B^\sigma +2 R^\rho{}_{(\mu} B_{\nu)} B_\rho +\frac12 g_{\mu\nu} \nabla_\rho\nabla_\sigma (B^\rho B^\sigma) \nn\\
&\quad& +\frac12 \Box (B_\mu B_\nu) - \nabla_\rho\nabla_{(\mu} (B_{\nu)} B^\rho) \bigg] -\frac{ \kappa}{2} \left[2 B_{\mu\lambda} B_\nu{}^\lambda -\frac12 g_{\mu\nu}B_{\rho\sigma}B^{\rho\sigma} \right] \nn\\
&\quad& +2 \kappa \left[\frac12 g_{\mu\nu} V - \frac{\partial V}{\partial X} B_\mu B_\nu \right] -\frac{1}{4}\left[2 F_{\mu\lambda} F_\nu{}^\lambda -\frac12 g_{\mu\nu}F_{\rho\sigma}F^{\rho\sigma} \right] \nn\\
&\quad& + \gamma_1 \left[ B_\rho B^\rho\left(2 F_{\mu\lambda} F_\nu{}^\lambda -\frac12 g_{\mu\nu}F_{\rho\sigma}F^{\rho\sigma} \right) + B_\mu B_\nu F_{\rho\sigma}F^{\rho\sigma}  \right] \nn\\
&\quad& + \gamma_2 \left[ -\frac12 g_{\mu\nu} B^\rho B_\lambda F_{\rho\sigma}F^{\lambda\sigma} + 2 B_\sigma B_{(\mu} F_{\nu)\rho}F^{\sigma\rho} + B^\rho B^\sigma F_{\mu\rho}F_{\nu\sigma} \right] =0  \,, \label{eom1} \\
E_{B}^\nu &\equiv& \gamma R^{\mu\nu}B_\mu +\kappa \nabla_\mu B^{\mu\nu} - 2\kappa \frac{\partial V}{\partial X} B^\nu  + \gamma_1 B^\nu F_{\rho\sigma}F^{\rho\sigma} + \gamma_2 B^\mu F_{\mu\lambda}F^{\nu\lambda} = 0 \,, \label{eom2} \\
E_{A}^\nu &\equiv& \nabla_\mu \left[ (1 - 4\gamma_1 B_\lambda B^\lambda) F^{\mu\nu} +4 \gamma_2 B_\lambda F^{\lambda [\mu} B^{\nu]} \right]= 0 \,, \label{eom3}
\eea
where $G_{\mu\nu} \equiv  R_{\mu\nu} - R g_{\mu\nu}/2$ is the Einstein tensor, $X = B_\mu B^\mu \pm b^2$, $\nabla_\mu$ is the covariant derivative, $\Box =\nabla_\mu \nabla^\mu$ is the d'Alembert operator. Parentheses $(\mu\nu)$ and square brackets $[\mu\nu]$ indicate symmetrization and antisymmetrization over the enclosed indices, respectively.

\section{Dyonic RN-like topological black holes in four dimensions} \label{solutiond4}

In this section, we construct the dyonic solution of \eqref{EBM}. A comparison of the coupling constants in the Maxwell sector of \eqref{action 2} shows that the bumblebee theory considered here differs from that studied in \cite{Liu:2024axg}. Moreover, we will illustrate the advantages of the theory developed in this work from the perspective of its dyonic solutions.

\subsection{Exact dyonic solution}
We now construct the static dyonic RN-like black hole solution with general topological horizons in $D=4$ dimensions. The most general static ansatz for the metric $g_{\mu\nu}$ and the Maxwell field potential $A_\mu$ can be written as
\bea
ds^2 &=& - h(r) dt^2 + \frac{dr^2}{f(r)}  + r^2 d\Omega_{2, k}^2  \,, \label{metric} \\
  A_{\1} &= & \phi (r) dt + p u d\varphi \,,  \label{maxwell}  
  \eea
where 
\be
d\Omega_{2, k}^2 = \frac{du^2}{1-k u^2} +(1-k u^2) d\varphi^2 \,, \label{omega2k}
\ee 
with $k = 1, 0, -1$, corresponding to the metric of the unit 2-sphere, the 2-torus or the unit hyperbolic 2-space, respectively. The constant $p$ denotes the magnetic charge parameter.
For the bumblebee field $B_\mu$, following Ref.~\cite{Casana:2017jkc}, we consider a  nonzero VEV $b_\mu$ oriented along a radial direction, taking the form
\be
B_{\1} = b_r (r) dr \,. \label{bumblebeefield} 
\ee
Imposing the condition $b_\mu b^\mu = b^2 = \textup{const.}$ leads to
\be
b_r = \frac{b}{\sqrt{f}} \,, \label{sol1}
\ee
which ensures that the bumblebee field strength vanishes, $b_{\mu\nu} = 0$, and that the potential, which may be chosen as $V=(B_\mu B^\mu - b^2)^2$, satisfies the vacuum conditions in Eqs.~(\ref{potential1})--(\ref{potential2}).
Next, substituting Eqs.~(\ref{metric})--(\ref{sol1}) into EOM~(\ref{eom3}), the $t$-component of $E_A^\mu$ yields
\be
\left(\frac{r^2 \sqrt{h}  }{\sqrt{f}} \ \frac{(1 -2 b^2 (2 \gamma_1+\gamma_2))f\phi'}{4 h}  \right)' = 0 \,,
\ee
where the prime denotes differentiation with respect to $r$. This equation can be integrated once to give
\be
\phi ' = \frac{q \sqrt{h}}{r^2 \sqrt{f} } \,, \label{result1}
\ee
where $q$ is an integration constant associated with the electric charge. 
Then, substituting Eqs.~(\ref{metric})--(\ref{sol1}) and (\ref{result1}) into EOMs~(\ref{eom1})--(\ref{eom2}), we obtain the nonzero components
\bea
E^t{}_t &\equiv& \frac{1 +b^2 \gamma}{r} \left[ f'+\frac{f}{r} + \frac{p^2 \left(1 -4 b^2 \gamma_1\right)+q^2 \left(1 -2 b^2 (2 \gamma_1+\gamma_2)\right)-4 k r^2}{4 r^3 \left(b^2 \gamma +1\right)} \right] = 0 \,, \label{ett} \\
E^r{}_r &\equiv& -\frac{b^2 \gamma f}{2 h} \Bigg[ h''-\frac{h'^2}{2 h}+ h' \left(\frac{f'}{2 f}-\frac{2 \left(b^2 \gamma +1\right)}{b^2 \gamma  r}\right) +\frac{2 h f'}{r f} -\frac{2 h \left(b^2 \gamma +1\right)}{b^2 \gamma  r^2} \nn\\
&\quad& +\frac{h \left(q^2 \left(6 b^2 (2 \gamma_1+\gamma_2)-1 \right)-p^2 \left(1+4 b^2 \gamma_1\right) +4 k r^2 \right)}{2 b^2 \gamma  r^4 f}  \Bigg] = 0 \,,  \label{err}\\
E^u{}_u &=& E^\varphi{}_\varphi  \equiv \frac{\left(b^2 \gamma +1\right) f }{2  h} \Bigg[h''-\frac{h'^2}{2 h} +\left(\frac{f'}{2 f}+\frac{1}{r}\right) h'+\frac{h f'}{r f} \nn \\
&\quad& + \frac{h \left(q^2 \left(2 b^2 (2 \gamma_1+\gamma_2)-1 \right)-p^2 \left(1 -4 b^2 \gamma_1\right)\right)}{2 r^4 \left(b^2 \gamma +1\right) f}\Bigg] = 0 \,,  \label{euu} \\
E^r_B &\equiv&  -\frac{b \gamma f^\frac{3}{2}}{ h} \left[ h'' - \frac{h'^2}{2 h} +\frac{f' h'}{2 f} +\frac{2 h f'}{r f} - \frac{2 h \left(2 \gamma_1 p^2-q^2 (2 \gamma_1+\gamma_2)\right)}{\gamma  r^4 f} \right] = 0 \,. \label{ebr}
\eea
From Eq.~(\ref{ett}), one can directly integrate to obtain
\be
f = \frac{1}{1 + b^2 \gamma } \Bigg[ k - \frac{m}{r}  +\frac{(1 -2 b^2 (2 \gamma_1+\gamma_2)) q^2}{4 r^2}  +\frac{\left(1 -4 b^2 \gamma_1\right) p^2}{4 r^2} \Bigg]  \,, \label{result3}
\ee
where $m$ is an integration constant related to the mass. Solving Eq.~(\ref{err}) for $h''$, and then substituting it together with Eq.~(\ref{result3}) into Eq.~(\ref{ebr}), we obtain
\be
h =  k - \frac{m}{r}  +\frac{(1 -2 b^2 (2 \gamma_1+\gamma_2)) q^2}{4 r^2}  +\frac{\left(1 -4 b^2 \gamma_1\right) p^2}{4 r^2}  \,. \label{result4}
\ee
Then, substituting Eqs.~(\ref{result3})--(\ref{result4}) back into Eq.~(\ref{err}), the existence of nonzero electric and magnetic charges $q$ and $p$ requires
\be
\gamma_1 = \frac{\gamma}{4 (2 +3 b^2 \gamma)} \,,\quad  \gamma_2 = -\frac{2 \gamma ( 1+ b^2 \gamma )}{(2 + b^2 \gamma) (2 +3 b^2 \gamma)} \,. \label{result5}
\ee
Finally, inserting Eqs.~(\ref{result3})–(\ref{result5}) into Eq.~(\ref{euu}), one finds that the equation is automatically satisfied. Combining with Eq.~(\ref{result1}) and rescaling $q\to q/\sqrt{1 + b^2 \gamma}$, the final solution reads
\bea
h &=& k - \frac{m}{r} + \frac{q^2 }{2 \left(2 + \ell \right) r^2 } + \frac{\left(1+\ell \right) p^2  }{2 \left(2+ 3\ell \right) r^2} \,, \nn\\ 
f &=& \frac{h}{1 + \ell} \,, \quad  \phi = \frac{q}{r} \,,
\eea
where the Lorentz violating parameter $\ell = b^2 \gamma$, and the couplings constants are 
\be
\gamma_1 = \frac{\gamma}{4 (2 +3\ell)} \,,\quad  \gamma_2 = -\frac{2 \gamma ( 1+\ell )}{(2 + \ell) (2 +3\ell)} \,.\label{result55}
\ee
Note that we have chosen the potential $\phi$ to vanish at infinity.
In the expression for the metric function $h$, the Lorentz-violating parameter $\ell$ does not appear explicitly in the mass term, since it can be absorbed into the mass integration constant $m$. However, once electromagnetic fields are coupled, this parameter can no longer be eliminated simultaneously from the metric function $h$ and $A_\mu$ through a redefinition of the electric and magnetic integration constants $q$ and $p$. More importantly, as in the neutral black-hole case, the Lorentz-violating parameter $\ell$ spoils the simple relation $f = h$. This reflects a fundamental feature of Einstein-bumblebee gravity: although imposing a vanishing potential fixes the functional form of the bumblebee vector $B_\mu$ and reduces the resulting gravitational field equations to second order, the theory remains intrinsically a higher-derivative gravity.
In the limit $\ell \to 0$, the solution reduces to the standard dyonic RN black hole with a generic topological horizon.  Alternatively, in the absence of both electric and magnetic charges with $k = 1$, the solution reduces to a Schwarzschild-like black hole, which has been extensively discussed in Ref.~\cite{Casana:2017jkc}.
The Kretschmann scalar, i.e., the square of the Riemann tensor, can be easily calculated as
\bea
R_{\mu\nu\rho\sigma} R^{\mu\nu\rho\sigma} &=& \frac{1}{r^8}\left[ \frac{14 p^4}{(3 \ell +2)^2}+\frac{28 p^2 q^2}{(\ell +1) (\ell +2) (3 \ell +2)}+\frac{14 q^4}{(\ell +1)^2 (\ell +2)^2} \right] \nn \\ 
&\quad & -\frac{m }{r^7} \left[\frac{24 p^2}{(\ell +1) (3 \ell +2)} +\frac{24 q^2}{(\ell +1)^2 (\ell +2)}\right] \nn \\
&\quad& +\frac{1}{r^6} \left[ -\frac{4 k p^2 \ell }{(\ell +1) (3 \ell +2)}-\frac{4 k q^2 \ell }{(\ell +1)^2 (\ell +2)}+\frac{12 m^2}{(\ell +1)^2} \right] \nn \\
&\quad&+\frac{8 k m \ell }{r^5 (\ell +1)^2} + \frac{4 k^2 \ell ^2}{r^4 (\ell +1)^2} \,,
\eea
which exhibits the similar singularity structure as the dyonic RN case, but clearly deviates from the corresponding Kretschmann invariant when $\ell \ne 0$. In particular, several terms contain $\ell $ as an overall prefactor in the numerator, with no counterparts in GR. These contributions vanish when $\ell=0$, confirming that they arise solely due to Lorentz-symmetry breaking.

\subsection{The difference from \cite{Liu:2024axg}
}
In \cite{Liu:2024axg}, the authors identified a set of static solutions in the purely electric case that differ from those obtained in this work. After choosing the coupling constants 
\bea
\gamma_1=-\frac{\gamma }{4 (\ell +2)}\,,\qquad\gamma_2 = 0\,,\label{gamma12 Liu}
\eea
 the metric $g_{\mu\nu}$, Maxwell field $A_{\1}$, and bumblebee field $B_{\1}$ take the following forms 
\bea
h &=& k - \frac{m}{r} + \frac{(1 + \ell)q^2 }{2 \left(2 + \ell \right) r^2 }  \,, \quad
f = \frac{h}{1 + \ell} \,, \nn \\
  \phi &=& \frac{\sqrt{\ell +1}q}{r}  \,, \quad b_r = \frac{b}{\sqrt{f}} \,.
\eea
Here, we correct a typo in the Maxwell field appearing in \cite{Liu:2024axg}. Because the coupling constant $\gamma_1$ in this solution differ from that considered in \eqref{result55}, the solution obtained in this work corresponds to a black hole in a distinct bumblebee theory.

Similarly, the above solution can be generalized to the dyonic case
\bea
A_{\1} &= & \phi (r) dt + p u d\varphi \,.
\eea
To obtain an analytical solution, we find that the function $h$ and the coupling constant $\gamma_2$ must be adjusted to 
\bea
h &=& k - \frac{m}{r} + \frac{(1 + \ell)q^2 }{2 \left(2 + \ell \right) r^2 }+\frac{ (\ell +1) (3 \ell +2)p^2}{2 (\ell +2)^2 r^2}\,,\nn\\
 \gamma_2&=&-\frac{2 \gamma   (\ell +1)}{ (\ell +2)^2}\frac{p^2}{q^2}\,.
\eea
We find that in this case, the coupling constant $\gamma_2$ depends on the integration constants $p$ and $q$. This indicates that the bumblebee theory \eqref{gamma12 Liu} considered in \cite{Liu:2024axg} cannot be directly extended to arbitrary dyonic black hole solutions, as the coupling constant would otherwise depend on the integration constants. Of course, for the special case $p = q$, such an extension remains possible.

Of course, we can also redefine the parameters $p \to pq$ to make 
\bea
\gamma_2=-\frac{2 \gamma   (\ell +1)}{ (\ell +2)^2} \,,
\eea
independent of the integration constants. However, in this case, the expressions for $A_{\1}$ and $h$ 
\bea
A_{\1} &= & \phi (r) dt + pq u d\varphi \,,\nn\\ 
h &=& k - \frac{m}{r} + \frac{(1 + \ell)q^2 }{2 \left(2 + \ell \right) r^2 }+\frac{ (\ell +1) (3 \ell +2)p^2q^2}{2 (\ell +2)^2 r^2}\,,
\eea
reveal that a purely magnetic black hole cannot be accommodated. Hence, in terms of the feasibility of extending to dyonic solutions, the theory considered here \eqref{result55} is better suited. In the following Sections \ref{taub nut dyonic} and \ref{highdim}, this approach can be further extended to Taub-NUT case and arbitrary even dimensions,  yielding consistent and well-defined dyonic solutions and coupling constants.

\section{Thermodynamics} \label{Thermodynamics}

Now we turn to the thermodynamics of the black hole. The event horizon, located at $r=r_h$, is determined by the condition $f(r_h)=0$, where the timelike Killing vector $\xi=\partial_t$ becomes null, i.e., $\xi_\mu\xi^\mu \big|_{r = r_h} = 0$. It is convenient to express the constant $m$ in terms of $r_h$, namely
\be
m = k r_h+\frac{p^2 (\ell +1)}{2 r_h (3 \ell +2)}+\frac{q^2}{2 r_h (\ell +2)} \,.
\ee
The black hole temperature $T$ is defined in terms of the surface gravity $K$, which is computed from the Killing vector $\xi$ as
\be
K^2 = - \frac{ \nabla^\mu \xi^\nu \nabla_\mu \xi_\nu }{2} \Big|_{r = r_h} \,, \quad  T = \frac{K}{2\pi} \,.\label{surface gravity and T}
\ee
Accordingly, the temperature reads
\be
 T = \frac{h'\left(r_h\right)}{4 \pi  \sqrt{\ell +1}} = \frac{k}{4 \pi  \sqrt{\ell +1} r_h} -\frac{q^2}{8 \pi  \sqrt{\ell +1} (\ell +2) r_h^3} -\frac{p^2 \sqrt{\ell +1}}{8 \pi  (3 \ell +2) r_h^3} \,.
\ee
In the limit $\ell \to 0 $, the temperature reduces to the GR result.
Next, in order to obtain the electric charge $Q_e$, it is convenient to introduce an antisymmetric tensor ${\cal F}^{\mu\nu}$ 
\be
{\cal F}^{\mu\nu} = (1 - 4\gamma_1 B_\lambda B^\lambda) F^{\mu\nu} +4 \gamma_2 B_\lambda F^{\lambda [\mu} B^{\nu]} \,,
\ee
to rewrite the Maxwell field EOM~(\ref{eom3}) in a compact form:
\be
 E_{A}^\nu \equiv \nabla_\mu {\cal F}^{\mu\nu} = 0  \,.
\ee
Equivalently, the above expression can be cast in the language of differential forms as
\bea
d*\mathcal{F}_{\2}=0\,,\quad *\mathcal{F}_{\2}=\frac{1}{2!2!}\epsilon_{\mu\nu\rho\sigma}\mathcal{F}^{\rho\sigma}dx^\mu\wedge dx^\nu\,, \label{eom3a}
\eea
where the subscript $(n)$ denotes an $n$-form, $d$ is the exterior derivative, $*$ the Hodge dual, and the Levi-Civita tensor $\epsilon_{\mu\nu\rho\sigma}$ is defined as $\epsilon_{\mu\nu\rho\sigma} = \sqrt{-g}\varepsilon_{\mu\nu\rho\sigma}$ with $\varepsilon_{0123}=1$.
In terms of the Eq.~(\ref{eom3a}), the electric charge $Q_e$ is then defined as
\bea
Q_e= -\frac{1}{2\kappa} \int_{\Omega_{2,k}} *\mathcal{F}_{\2} =\frac{1}{2\kappa}  \int_{\Omega_{2,k}} \sqrt{-g}\varepsilon_{tru\varphi} \mathcal{F}^{rt} du d\varphi = \frac{q w_2 \sqrt{\ell +1}}{\kappa (\ell +2)}\,,\label{compute Qe}
\eea
where $w_2 = \int_{\Omega_{2,k}} d u d \varphi$. For $k =1$, corresponding to the unit 2-sphere, one has $w_2 = 4\pi$.
The electric potential $\Phi_e$ is defined through $A_{\1}$
\bea
\Phi_e=i_\xi A_{\1}|^{r=r_h}_{r\rightarrow\infty} = \xi^\mu A_\mu |^{r=r_h}_{r\rightarrow\infty} =\frac{q}{r_h}\,,\label{electric potential}
\eea
where $i_\xi$ denotes a contraction of $\xi^\mu$ on the index of the $1$-form $A_{\1}$. 
The Maxwell field strength $F_{\2} = d A_{\1}$ satisfies the Bianchi identity $dF_{\2}=0$, which allows the magnetic charge $Q_m$ to be defined in terms of it
\bea
Q_m=\frac{1}{2\kappa} \int_{\Omega_{2,k}} F_{\2} = \frac{1}{2\kappa}  \int_{\Omega_{2,k}} F_{u\varphi} du d\varphi=\frac{w_2 p}{2\kappa} \,.\label{magnetic charge}
\eea
According to \cite{Rasheed:1997ns}, since $*\mathcal{F}_{\2}$ is closed when the on-shell condition is satisfied (i.e. the EOMs are satisfied), we can define a corresponding 1-form $\mathcal{A}_{\1}$ satisfying $*\mathcal{F}_{\2} = d \mathcal{A}_{\1} $. The $\mathcal{A}_{\1}$ is given by
\be
\mathcal{A}_{\1} = \frac{2 p (\ell +1)^{3/2}}{ r (3 \ell +2)} dt -\frac{2 q \sqrt{\ell +1}}{ (\ell +2)} u d\varphi \,.\label{dual 1 form}
\ee
Similarly, the magnetic potential $\Phi_m$ can therefore be defined in terms of $\mathcal{A}_{\1}$
\bea
\Phi_m=i_\xi \mathcal{A}_{\1}|^{r=r_h}_{r\rightarrow\infty} = \xi^\mu \mathcal{A}_\mu |^{r=r_h}_{r\rightarrow\infty}  =\frac{2 p (\ell +1)^{3/2}}{ (3 \ell +2) r_h}\,.\label{magnetic potential}
\eea

To complete the thermodynamic quantities, we now turn to the mass and entropy of the black hole. The Komar mass associated with the timelike Killing vector $\xi$ is
\bea
M_\textup{K} = - \frac{1}{\kappa} \int_{\Omega_{2,k} } * d\xi\big|_{r\to\infty}
 = \frac{2}{\kappa} \int_{\Omega_{2,k} } du d\varphi\sqrt{-g}\varepsilon_{tru\varphi} \nabla^r \xi^{t} \big|_{r\to\infty} = \frac{ m w_2 }{\kappa \sqrt{1+\ell}}   \,. \label{komar}
\eea
 It is evident that the integration constant $m$ is proportional to the Komar mass. One may further verify that the same integration constant is also proportional to the Arnowitt-Deser-Misner mass. 
The Wald entropy formula~\cite{Wald:1993nt,Iyer:1994ys} reads
\be
S_\textup{W} = -\frac{\pi }{\kappa} \int_{\Omega_{2,k}} r^2 \frac{\partial L_1}{\partial R_{\mu\nu\rho\sigma}}\epsilon_{\mu\nu}\epsilon_{\rho\sigma} \bigg|_{r\to r_h}= \frac{w_2 \pi}{\kappa}   (\ell +2) r_h^2 \,. \label{waldentropy}
\ee
where the binormal is $\epsilon_{\mu\nu} = \sqrt{h/f} (\delta_\mu^t \delta_\nu^r -\delta_\mu^r \delta_\nu^t ) $. However, the Komar mass and Wald entropy, together with the temperature, electric and magnetic charges, and their corresponding potentials, do not satisfy the first law of black hole thermodynamics:
\bea
\delta M_\textup{K}  \neq T\delta S_\textup{W} +\Phi_e\delta Q_e+\Phi_m\delta Q_m\,.
\eea
Similar discrepancies have also been observed in the neutral cases, as discussed in Refs.~\cite{An:2024fzf,Chen:2025ypx}.
In fact, similar violations of the first law due to naive application of definitions have also been reported in other theories, such as four-dimensional gauged supergravities~\cite{Lu:2013ura,Chow:2013gba,Wu:2015ska}, Horndeski gravity~\cite{Feng:2015oea,Feng:2015wvb}, and related setups~\cite{Ma:2022nwq,Liu:2022wku}. Motivated by this, we employ the Wald formalism~\cite{Wald:1993nt,Iyer:1994ys} to consistently recompute the mass and entropy.

\subsection{Wald formalism} \label{Wald}

In 1993, Wald developed the covariant phase space formalism \cite{Wald:1993nt,Iyer:1994ys}. Within this framework, he demonstrated that in a diffeomorphism-invariant theory of gravity, the Killing vector \(\xi^\mu\) generates all conserved charges, thereby unifying these quantities into a single differential relation, i.e., the first law of black hole mechanics. In other words, the first law emerges as a natural consequence of Noether’s theorem, ensuring that any diffeomorphism-invariant theory of gravity necessarily admits a self-consistent black hole thermodynamics.

A direct consequence of the Wald formalism is the Wald entropy formula, which extends the Bekenstein-Hawking entropy and possesses remarkable universality. In higher-derivative gravity theories, black hole entropy is no longer proportional to the horizon area. The Wald entropy formula, however, correctly accounts for these corrections \cite{Fan:2014ala,Ma:2023qqj,Hu:2023gru} and is thus widely used in many black hole–related effective theories \cite{Liu:2014dva,Ma:2022nwq,Liu:2022wku,Ma:2024ynp,Guo:2025muo,Guo:2025ohn,Chen:2025ary}.
 Thus far, the Wald formalism and the Wald entropy formula have been applied to a broad range of extended gravity theories, including tensor-scalar theories \cite{Liu:2013gja,Lu:2014maa,Feng:2015oea,Feng:2015wvb}, higher-derivative graivty theories coupled maxwell
field \cite{Fan:2014ixa,Feng:2015sbw}, Maxwell theory and its extensions \cite{Gao:2003ys,Li:2016nll,Liu:2014tra,Gao:2003ys,Fan:2014ala,Lu:2013ura,Chow:2013gba,Fan:2017bka}, as well as the neutral bumblebee theory \cite{An:2024fzf,Chen:2025ypx}.

In fact, the Wald entropy formula is simply a corollary of the Wald formalism and can be obtained from it only under specific conditions. Indeed, previous studies of Horndeski theory and bumblebee theory have demonstrated that the Wald entropy formula fails to provide a black hole entropy consistent with the first law of thermodynamics \cite{An:2024fzf,Chen:2025ypx,Feng:2015oea,Feng:2015wvb}. More precisely, the dependence of the Noether charge on the Killing vector involves only \(\xi^\mu\) and \(\nabla^\mu \xi^\nu\). On the black hole horizon, if all fields are smooth, one has \(\xi \to 0\) and \(\nabla^\mu \xi^\nu \to K \epsilon^{\mu\nu}\), from which the Wald entropy formula follows. However, the \(B_\mu\) field (\ref{bumblebeefield}, \ref{sol1}) diverges on the black hole horizon, rendering the Wald entropy formula inapplicable. Although this divergence---like that of $g_{rr}$---is purely a coordinate artifact, the complete thermodynamic quantities must nevertheless be derived directly from the original Wald formalism.

Consider a diffeomorphism invariant theory whose action ${\cal S}$ in the language of differential form is given by
\be
{\cal S}[\psi] =\frac{1}{2\kappa} \int \mathbf{L}(\psi) \,,
\ee
where $\psi$ collectively denotes all dynamical fields in the system. The Lagrangian $D$-form $\mathbf{L}$ is the Hodge dual of the Lagrangian density $ L =  L_1 +L_2$, namely,
\be
\mathbf{L} = * L = \frac{\sqrt{-g}}{D!}\varepsilon_{\alpha_1\alpha_2\dots\alpha_D} L  dx^{\alpha_1} \wedge dx^{\alpha_2} \wedge \dots \wedge dx^{\alpha_D} \,.
\ee
The variation of the Lagrangian $\mathbf{L}$ takes the general form
\be
\delta \mathbf{L} = \mathbf{E}_\psi \delta \psi + d \mathbf{\Theta} (\psi, \delta \psi) \,,
\ee
where  $\mathbf{E}_\psi$ represents the EOM, and $\mathbf{\Theta}$ is the symplectic potential $(D-1)$-form. Explicitly, 
\bea
 \mathbf{\Theta} &=& \frac{\sqrt{-g}}{(D-1)!}\varepsilon_{\mu\alpha_1\alpha_2\dots \alpha_{D-1}} \Theta^\mu  dx^{\alpha_1} \wedge dx^{\alpha_2} \wedge \dots \wedge dx^{\alpha_{D-1}} \,,\\
 i_\xi \mathbf{\Theta} &=& \frac{\sqrt{-g}}{(D-2)!}\varepsilon_{\mu\nu\alpha_1\alpha_2\dots \alpha_{D-2}} \Theta^\mu \xi^\nu dx^{\alpha_1} \wedge dx^{\alpha_2} \wedge \dots \wedge dx^{\alpha_{D-2}} \,.
\eea
From $ \mathbf{\Theta}$, one can define the symplectic current $(D-1)$-form $\mathbf{\omega}$ as
\be
\mathbf{\omega}  (\psi, \delta_1 \psi, \delta_2 \psi) = \delta_1 \mathbf{\Theta}  (\psi, \delta_2 \psi) -\delta_2 \mathbf{\Theta}  (\psi, \delta_1 \psi) \,. \label{omega}
\ee
We next specialize to a variation induced by an infinitesimal diffeomorphism generated by an arbitrary vector field $\xi^\mu$, the dynamical fields transform as
\be
\delta_\xi \psi = \mathscr{L}_\xi \psi \,,
\ee
with $\mathscr{L}_\xi$ representing the Lie derivative along $\xi^\mu$. The corresponding variation of the Lagrangian $\mathbf{L}$ can likewise be expressed in two equivalent forms. On the one hand, from the general variation formula we have
\be
\delta_\xi \mathbf{L} =  \mathbf{E}_\psi \delta_\xi \psi + d \mathbf{\Theta} (\psi, \delta_\xi \psi) \,, \label{deltaL1}
\ee
while on the other hand, using the diffeomorphism invariance of the theory, it is given by the Lie derivative of $\mathbf{L}$, namely
\be
\mathscr{L}_\xi \mathbf{L} = i_\xi d\mathbf{L} + d (i_\xi \mathbf{L}) = d (i_\xi \mathbf{L})\,,\label{deltaL2}
\ee
where we have used $d\mathbf{L} = 0$ since the $\mathbf{L}$ is a $D$-form on a $D$-dimensional manifold. One can introduce a Noether current $(D-1)$-form $\mathbf{J}_\xi$, defined by
\be
\mathbf{J}_\xi = \mathbf{\Theta}  (\psi, \delta_\xi \psi) - i_\xi \mathbf{L} \,.
\ee
Combining Eqs.~(\ref{deltaL1}) and (\ref{deltaL2}) gives 
\be
d \mathbf{J}_\xi  = -\mathbf{E}_\psi \delta_\xi \psi  \,,
\ee
which vanishes on-shell. Consequently, there exists a Noether charge $(D-2)$-form $\mathbf{Q}_\xi$ such that
\be
\mathbf{J}_\xi = d \mathbf{Q}_\xi \,.
\ee
The Noether charge can be expressed as
\be
\mathbf{Q}_\xi = \frac{\sqrt{-g}}{(D-2)! 2!}\varepsilon_{\mu \nu\alpha_1\alpha_2\dots \alpha_{D-2}  } Q_\xi^{\mu\nu}  dx^{\alpha_1} \wedge dx^{\alpha_2} \wedge \dots \wedge dx^{\alpha_{D-2}}  \,.
\ee
Varying the Noether current $\mathbf{J}_\xi$, we obtain
\bea
\delta \mathbf{J}_\xi &=& \delta \mathbf{\Theta}  (\psi, \delta_\xi \psi) - i_\xi \delta\mathbf{L} \nn \\
&=&  \delta \mathbf{\Theta}  (\psi, \delta_\xi \psi) - i_\xi d  \mathbf{\Theta}  (\psi, \delta \psi) - i_\xi (\mathbf{E}_\psi \delta \psi) \nn \\
&=&  \delta \mathbf{\Theta}  (\psi, \mathscr{L}_\xi \psi) - \mathscr{L}_\xi \mathbf{\Theta}  (\psi, \delta \psi) +d (i_\xi \mathbf{\Theta}(\psi, \delta \psi) )  \,. \label{deltaJxi}
\eea
where we have used the on-shell condition, $\mathbf{E}_\psi = 0$.
According to Eq.~(\ref{omega}), the first two terms of the right hand side of the Eq.~(\ref{deltaJxi}) can be identified as the symplectic current $(D-1)$-form $\mathbf{\omega}$:
\bea
\mathbf{\omega}(\psi, \delta \psi, \mathscr{L}_\xi \psi) &=& \delta \mathbf{\Theta}  (\psi, \mathscr{L}_\xi \psi) - \mathscr{L}_\xi \mathbf{\Theta}  (\psi, \delta \psi) \nn \\
&=& \delta \mathbf{J}_\xi -d (i_\xi \mathbf{\Theta}(\psi, \delta \psi) ) \nn  \\
&=& d (\delta \mathbf{Q}_\xi -i_\xi \mathbf{\Theta}(\psi, \delta \psi) ) \,.
\eea
To make contact with the first law of black hole thermodynamics, we take $\xi^\mu$ to be the timelike Killing vector that becomes null on the horizon. Wald shows that the variation of the Hamiltonian $\mathcal{H}$ with respect to the integration constants of a specific solution $\psi$ is given by
\be
\delta {\cal H} = \frac{1}{2\kappa} \int_C \mathbf{\omega} (\psi, \delta \psi, \mathscr{L}_\xi \psi) = \frac{1}{2\kappa} \int_{\Omega_{D-2}} ( \delta \mathbf{Q}_\xi -i_\xi \mathbf{\Theta}(\psi, \delta \psi)) \,,\label{Wald formalism H}
\ee
where $C$ denotes a Cauchy surface, and $\Omega_{D-2}$ is its boundary, which has two components, one at infinity and one on the horizon. Thus according to the Wald formalism, the first
law of black hole thermodynamics is a consequence of
\be
\delta {\cal H}_{\infty} =\delta {\cal H}_{r_h} \,.
\ee

For $D= 4$ dimensional EbM theory with $L = L_1 + L_2$, we have
\bea
\Theta^\mu = \Theta^\mu_1  + \Theta^\mu_2  \,, \label{surfaceterms}
\eea
with
\bea
\Theta^\mu_1 &=&  g^{\mu\rho}g^{\nu\sigma} (\nabla_\sigma \delta g_{\nu\rho} - \nabla_\rho \delta g_{\nu\sigma} ) + \frac{\gamma}{2} \Big[ 2 g^{\mu\lambda} B^\rho B^\sigma \nabla_\rho \delta g_{\lambda\sigma}  \nn \\
&\quad& - g^{\mu\lambda} B^\rho B^\sigma \nabla_\lambda \delta g_{\rho\sigma} - g^{\rho\lambda} B^\mu B^\sigma \nabla_\sigma \delta g_{\rho\lambda} - 2 g^{\rho\lambda} \nabla_\rho (B^\mu B^\sigma)\delta g_{\sigma\lambda}  \nn \\
&\quad& + g^{\mu\lambda} \nabla_\lambda (B^\rho B^\sigma)\delta g_{\rho\sigma}  + g^{\rho\lambda} \nabla_\sigma (B^\mu B^\sigma)\delta g_{\rho\lambda}  \Big] - 2\kappa B^{\mu\nu} \delta B_\nu  \,,  \label{surfaceterm1}\\
\Theta^\mu_2 &=& -  \left[ (1 - 4 \gamma_1 B^\lambda B_\lambda) F^{\mu\nu} + 4 \gamma_2 F^{\lambda [\mu} B^{\nu]} B_\lambda \right] \delta A_\nu = - {\cal F}^{\mu\nu} \delta A_\nu \,.  \label{surfaceterm3}
\eea
Specializing to a variation induced by $\xi^\mu$, and substituting the following equations
\bea
\delta_\xi g_{\mu\nu} &=& \nabla_\mu \xi_\nu + \nabla_\nu \xi_\mu \,, \\
\delta_\xi B_\mu &=& \xi^\rho \nabla_\rho B_\mu + B_\rho \nabla_\mu \xi^\rho \,, \\
\delta_\xi A_\mu &=& \xi^\rho \nabla_\rho A_\mu + A_\rho \nabla_\mu \xi^\rho \,, 
\eea
into Eq.~(\ref{surfaceterms}), the Noether charge $Q_\xi^{\mu\nu} = Q_1^{\mu\nu} +Q_2^{\mu\nu}$ can be obtained from 
\be
\nabla_\nu Q_\xi^{\mu\nu} = J_\xi^\mu= \Theta^\mu(\delta_\xi) - \xi^\mu L +\textup{EOM} \,,
\ee
and is given by
\bea
Q_1^{\mu\nu} &=& -2 \nabla^{[\mu} \xi^{\nu]} + 2 \gamma \left[\xi_\lambda \nabla^{[\mu} (B^{\nu]} B^\lambda) -\xi^{[\mu} \nabla_{\lambda} (B^{\nu]} B^\lambda) -B^\lambda B^{[\mu} \nabla_{\lambda} \xi^{\nu]} \right] -2\kappa B^{\mu\nu} B^\lambda \xi_\lambda \,,\\ 
Q_2^{\mu\nu} &=& - {\cal F}^{\mu\nu} A^\lambda \xi_\lambda  \,.\label{Noether charge}
\eea
To specialize our black hole ansatz (\ref{metric})--(\ref{sol1}), the result for the gravity part is well established and is given by
\bea
\mathbf{Q}_{\xi1} &=& r^2 \left[\frac{h' \sqrt{f}}{ \sqrt{h}} +\frac{\ell  \sqrt{f} \left(r h'-4 h\right)}{2 r \sqrt{h}} \right] du \wedge d\varphi \,,\\
i_{\xi}  \mathbf{\Theta}_1 &=& r^2\Bigg[ \left(\frac{2 \delta f \sqrt{h}}{r \sqrt{f}} +\frac{\delta h' \sqrt{f}}{ \sqrt{h}} - \frac{ h' \sqrt{f} \delta h}{2 h^{3/2}} + \frac{ h' \delta f}{2\sqrt{f h}} \right) \nn \\
&\quad&  +\ell \left(\frac{\delta f \sqrt{h}}{r \sqrt{f}} -\frac{\delta h \sqrt{f}}{r \sqrt{h}} +\frac{\delta f h'}{4 \sqrt{hf} } + \frac{\sqrt{f} \delta h'}{2 \sqrt{h}}-\frac{\delta h \sqrt{f} h'}{4 h^{3/2}}\right) \Bigg] du \wedge d\varphi \,, \\
\delta\mathbf{Q}_{\xi1} -i_{\xi} \mathbf{\Theta}_1 &=&  r^2 \Bigg[ - \frac{2 \delta f \sqrt{h}}{r \sqrt{f}} (1 +\ell )\Bigg] du \wedge d\varphi \,.
\eea
These results are consistent with previous works~\cite{An:2024fzf,Chen:2025ypx}. For the electromagnetic part, it is convenient to introduce a total derivative term $d (\Psi \delta A_{\1})$~\cite{Lu:2013ura,Ma:2022nwq,Kim:2023ncn} to the integrand of $\delta {\cal H}$. This term does not affect the final result, since $\delta\mathbf{Q}_{\xi } - i_\xi \mathbf{\Theta}$ is closed, and is introduced to circumvent the Dirac string singularity that would otherwise appear in $\delta \mathcal{H}$ due to the presence of a magnetic charge. The scalar $\Psi$ is defined via 
\be
d\Psi = i_\xi * \mathcal{F}_{\2} \quad \Rightarrow \quad \Psi = -i_\xi \mathcal{A}_{\1} = -\frac{2 p (\ell +1)^{3/2}}{ (3 \ell +2) r} \,.\label{scalar Psi}
\ee
Thus we have
\bea
&&\delta\mathbf{Q}_{\xi2} = \delta (-*\mathcal{F}_{\2} i_\xi A_{\1}) = -\delta (*\mathcal{F}_{\2}) i_\xi A_{\1} -*\mathcal{F}_{\2} \delta ( i_\xi A_{\1}) \,,\\
&&i_{\xi} \mathbf{\Theta}_2 = i_\xi (- *\mathcal{F}_{\2} \wedge \delta A_{\1}) = -i_\xi (*\mathcal{F}_{\2}) \wedge \delta A_{\1} -*\mathcal{F}_{\2} \delta ( i_\xi A_{\1}) \,,\\
&&d(\Psi  \delta A_{\1}) = d\Psi  \delta A_{\1} + \Psi  \delta dA_{\1} = i_\xi (*\mathcal{F}_{\2}) \wedge \delta A_{\1} - i_\xi \mathcal{A}_{\1} \delta F_{\2} \,,\\
&&\delta\mathbf{Q}_{\xi2} -i_{\xi} \mathbf{\Theta}_2 - d (\Psi \delta A_{\1}) =  - i_\xi A_{\1} \delta (*\mathcal{F}_{\2}) + i_\xi \mathcal{A}_{\1} \delta F_{\2} \,.
\eea
Finally, we have
\be
\delta {\cal H} = \frac{w_2 }{\kappa} \sqrt{1 +\ell} \delta m \,,
\ee
At infinity, we have
\be
\delta {\cal H}_{\infty} = \delta M = \frac{w_2 }{\kappa}  \sqrt{1 +\ell} \delta m \,,
\ee
which indicates that the mass $M$ is 
\be
M = \frac{m w_2 }{\kappa} \sqrt{1 +\ell} \,, \label{noethercharge}
\ee
In the limit $\ell \to 0 $, the Noether mass~(\ref{noethercharge}) coincides with the Komar mass~(\ref{komar}), reproducing the standard GR result as expected.
At horizon, we have
\be
\delta {\cal H}_{r_h} = T \delta S + \Phi_e \delta Q_e  + \Phi_m \delta Q_m \,,
\ee
and the entropy $S$ can be easily calculated by
\be
S = \frac{2\pi w_2  r_h^2 (1+\ell)}{\kappa} \,. \label{entropynoether}
\ee
In the limit $\ell \to 0 $, the entropy~(\ref{entropynoether}) reduce to the area entropy in GR.
Thus the first law of black hole thermodynamics reads
\be
\delta M = T \delta S +\Phi_e \delta Q_e +\Phi_m \delta Q_m \,. \label{1stlaw}
\ee
The integral first law of black hole thermodynamics, also called Smarr formula, is given by
\be
M = 2 T S  +\Phi_e Q_e +\Phi_m Q_m \,. \label{1stlaw2}
\ee
When the topological parameter $k= 0$, corresponding to planar black holes, there exists an additional generalized Smarr relation~\cite{Liu:2015tqa}
\be
M = \frac23 (T S  +\Phi_e Q_e +\Phi_m Q_m) \,. \label{1stlaw3}
\ee
It is evident that, although the Lorentz-violating parameter $\ell$ affects nearly all thermodynamic quantities of the dyonic RN-like black hole---except for the electric potential and magnetic charge---all forms of the first law of black hole thermodynamics, Eqs.~(\ref{1stlaw})--(\ref{1stlaw3}), remain consistent with their GR counterparts. Finally, since all relevant physical quantities have now been determined, it is convenient to rewrite the black-hole solution directly in terms of these physical quantities rather than in terms of the integration constants. The metric, electricmagnetic field, and bumblebee field can be expressed as
\bea
ds^2 &=& \left[ k - \frac{\kappa M}{r w_{2,k} \sqrt{1+\ell}} + \frac{\kappa^2 Q_e^2 (2+\ell) }{2 w_{2,k}^2 r^2 } + \frac{2 \kappa^2 \left(1+\ell \right) Q_m^2  }{ w_{2,k}^2 \left(2+ 3\ell \right) r^2} \right] dt^2 \nn \\
&\quad& + \frac{1+ \ell}{ k - \frac{\kappa M}{r w_{2,k} \sqrt{1+\ell}} + \frac{\kappa^2 Q_e^2 (2+\ell) }{2 w_{2,k}^2 r^2 } + \frac{2 \kappa^2 \left(1+\ell \right) Q_m^2  }{ w_{2,k}^2 \left(2+ 3\ell \right) r^2} } dr^2 + r^2 \left(\frac{du^2}{1-k u^2} +(1-k u^2) d\varphi^2 \right) \,,  \nn\\
A &=& \frac{\kappa (2+\ell) Q_e }{ r w_{2,k} \sqrt{1+\ell}  } dt + \frac{2 \kappa Q_m u  }{ w_{2,k}} d\varphi \,,  \quad
B = \frac{b }{ \sqrt{k - \frac{\kappa M}{r w_{2,k} \sqrt{1+\ell}} + \frac{\kappa^2 Q_e^2 (2+\ell) }{2 w_{2,k}^2 r^2 } + \frac{2 \kappa^2 \left(1+\ell \right) Q_m^2  }{ w_{2,k}^2 \left(2+ 3\ell \right) r^2}} } \,,  \nn\\
\gamma_1 &=& \frac{\gamma}{4 (2 +3\ell)} \,,\quad  \gamma_2 = -\frac{2 \gamma ( 1+\ell )}{(2 + \ell) (2 +3\ell)} \,,\quad \ell = b^2 \gamma \,.
\eea
For the spherical topological case, i.e. $k = 1$, and $w_{2,k} = 4\pi$, the dyonic RN-like black hole possesses two horizons,
\be
r_{\pm} = \frac{M \pm \sqrt{M^2-2 (\ell +1) \left((\ell +2) Q_e^2+\frac{4 (\ell +1) Q_m^2}{3 \ell +2}\right)}}{\sqrt{\ell +1}} \,,
\ee
where the mass $M$, electric charge $Q_e$, and magnetic charge $Q_m$ must satisfy the following condition:
\be
M^2 \ge 2 (\ell +1) \left((\ell +2) Q_e^2+\frac{4 (\ell +1) Q_m^2}{3 \ell +2}\right) \,.
\ee
In the limit $Q_e \to 0 $ or $Q_m \to 0 $, the solution reduces to purely electric or purely magnetic black hole, respectively. In the limit $\ell \to 0 $, the above solution smoothly reduces to the standard dyonic RN black hole in GR, as expected.

\section{Dyonic Taub-NUT-like black holes } \label{taub nut dyonic}
Owing to the Misner string singularity, the Taub–NUT solution \cite{Taub:1950ez,Newman:1963yy} constitutes a nontrivial generalization of conventional static black holes.

\subsection{Exact solution}

The bumblebee theory we consider  \eqref{result55} admits a Taub-NUT dyonic solution 
\bea 
ds^2&=&-h(dt+2N u d\varphi)^2+\frac{dr^2}{f}+(r^2+N^2)d\Omega_{2, k}^2\,, \nn\\
A_{\1}&=&\sqrt{\frac{3 \ell +2}{(\ell +1) (\ell +2)}}\Big[
\phi (dt+2N u d\varphi)+p u d\varphi
\Big]+c_edt\,,\nn \\
B_{\1}&=&b_r dr \,,
\eea 
where $d\Omega_{2, k}$ is given by (\ref{omega2k}), $N$ is called NUT parameter, and
\bea 
h(r)&=&\frac{(r^2-N^2) k }{r^2+N^2} - \frac{ m r}{r^2+N^2} + \frac{p^2+q^2}{2 (\ell +2) (r^2+N^2)} \,, \nn \\  
\phi(r)&=&\frac{1}{2N} \left(q \sin X(r) +p \cos X(r) -p \right)\,,\nn\\ 
b_r(r)&=&\frac{b}{\sqrt{f}}\,,\qquad f(r)=\frac{h(r)}{1+\ell}\,, \quad \ell = b^2 \gamma \,, \nn \\
X(r) &=& 2\sqrt{\frac{(\ell +1) (\ell +2)}{3 \ell +2}}\tan ^{-1}\big(\frac{N}{r}\big) \,.\label{NUT solution}
\eea 
For $p=q=0$, and $k=1$, the solution reduces to that  given in Ref.~\cite{Chen:2025ypx}. For $N=0$, the solution reduces to that presented in Section~\ref{solutiond4}. For convenience, we redefine the magnetic charge parameter \(p\), which differs from the one in \eqref{maxwell} by a constant factor $\sqrt{(3 \ell +2)/((\ell +1) (\ell +2))}$. The parameter \(c_e\) is a gauge constant, which does not affect the solution. In previous RN-like solution, it was set to zero.

Unlike conventional black holes, the Taub-NUT solution involves a Misner string singularity, and its thermodynamic analysis remains debated. In this work, we adopt the approach of \cite{Liu:2022wku} to evaluate the thermodynamics. We begin with the thermodynamic quantities that are well-defined, and then proceed to evaluate those requiring special treatment. In order to compute a well-define NUT potential, we can only handle the sphere topology $k=1$ case.

\subsection{Normal thermodynamic quantities}

The temperature $T$ of the Taub–NUT black hole is still determined by the surface gravity $K$, as expressed in \eqref{surface gravity and T}
\bea
T=\frac{1}{4 \pi r_h \sqrt{\ell +1}}\left[1-\frac{p^2+q^2}{2 (\ell +2) (r_h^2+N^2)}\right]\,.
\eea 

Using the expression for the Maxwell field \(A_{\1}\) and its EOM, one can determine the electric potential \(\Phi_e\) and the electric charge \(Q_e\). Likewise, the magnetic charge \(Q_m\)  can be obtained via the Bianchi identity $dF_{\2}=0$, and the magnetic potential \(\Phi_m\) can be calculated by introducing the dual 1-form field \(\mathcal{A}_{\1}\) satisfying $*\mathcal{F}_{\2} = d \mathcal{A}_{\1} $, which is given by
\be 
\mathcal{A}_{\1}=\mathcal{A}_{t}dt+\mathcal{A}_{\varphi}d\varphi+c_m dt\,,  \label{dualfield}
\ee
with
\bea 
\mathcal{A}_{t}&=&-\frac{\sqrt{\ell +1}}{  N (\ell +2)}(q \cos X -p \sin X)  \,,\cr 
\mathcal{A}_{\varphi}&=&-\frac{2 u \sqrt{\ell +1}}{\ell +2}(q \cos X -p \sin X) \,.
\eea
In the limit $N \to 0$, Eq.~(\ref{dualfield}) reduces to the dyonic RN-like case (\ref{dual 1 form}), with the gauge parameter $c_m = 0$ and $p\to p\sqrt{((\ell +1) (\ell +2))/(3 \ell +2)}$.
The electric and magnetic potentials $\Phi_e$ and $\Phi_m$ are given by
\bea 
\Phi_e&=&\frac{1 }{2 N}\sqrt{\frac{3 \ell +2}{(\ell +1) (\ell +2)}} \Big[q \sin X(r_h) +p \cos X(r_h) -p\Big] \,,\label{electricp} \\ 
\Phi_m&=&\frac{\sqrt{\ell +1}}{  N (\ell +2)}\Big[ q-q \cos X(r_h) +p \sin X(r_h)\Big]\,. \label{mageneticp} 
\eea 
Similarly, in the $N \to 0$ limit, Eqs.~\eqref{electricp}--\eqref{mageneticp} reduce to Eqs.~\eqref{electric potential} and \eqref{magnetic potential}.

Because of the Misner string, computing the electric charge requires careful treatment of the integration. We compute the electric charge \(Q_e\) following the same procedure as in \eqref{compute Qe}. Unlike in the static black hole case ($N=0$), the presence of the NUT parameter introduces a \( dr \wedge d\varphi \) component in the expression for \( *\mathcal{F}_{\2} \)
\bea
*\mathcal{F}_{\2}&=&\mathcal{Q}_{tr}dt\wedge dr+\mathcal{Q}_{r\varphi}dr\wedge d\varphi+\mathcal{Q}_{u\varphi}du\wedge d\varphi \,,\nn\\
\mathcal{Q}_{r\varphi}&=&-2Nu\mathcal{Q}_{tr}\,, \nn\\
\mathcal{Q}_{tr}&=&\frac{2 (\ell +1) (p +2 N \phi)}{\sqrt{(\ell +2) (3 \ell +2)} (r^2+N^2)}
\,, \qquad 
\mathcal{Q}_{u\varphi}=\frac{2 \sqrt{3 \ell +2} (r^2 +N^2)\phi'}{ (\ell +2)^{3/2}}\,.\label{nut star cal F}
\eea 
Following the integration procedure outlined in \cite{Liu:2022wku}, the electric charge \(Q_e\) 
\bea 
Q_e
&=&-\frac{1}{2\kappa}\int d\varphi\Bigg[ \int_{-1}^1\mathcal{Q}_{u\varphi}(r\to\infty,u)du - \int_{r_h}^{\infty}\mathcal{Q}_{r\varphi}(r,u)\Big|_{u=-1}^{u=1}dr\Bigg] \nn\\
&=&-\frac{w_2}{4\kappa}\mathcal{A}_\varphi|_{u=-1}^{u=1} =\frac{w_2 \sqrt{\ell +1}}{\kappa  (\ell +2)}\big[q \cos X(r_h) -p \sin X(r_h)\big]  \,,\label{electric charge BB}
\eea
can be determined.  Here, the charge integral is evaluated over \( S^2 \) ($w_2 = 4\pi$); therefore, the \( dt \wedge dr \) component does not contribute to it. In the next subsection, it will be shown that this component corresponds to the charge induced by the NUT parameter.

As in the case of the electric charge, we compute the magnetic charge \(Q_m\) using the procedure described in \eqref{magnetic charge}. Here
\bea 
F_{\2}&=&\mathcal{P}_{tr}dt\wedge dr+\mathcal{P}_{r\varphi}dr\wedge d\varphi+\mathcal{P}_{u\varphi}du\wedge d\varphi\,,\nn\\
\mathcal{P}_{r\varphi}&=&-2Nu\mathcal{P}_{tr}\,,\nn\\
\mathcal{P}_{tr}&=& -\sqrt{\frac{3 \ell +2}{(\ell +1) (\ell +2)}}\phi'\,,\qquad
\mathcal{P}_{u\varphi}=\sqrt{\frac{3 \ell +2}{(\ell +1) (\ell +2)}}\big(p+2N\phi\big)\,.\label{nut F}
\eea 
We continue to employ the same integration method as for the electric charge, allowing the magnetic charge to be computed straightforwardly 
\bea 
Q_m
&=&\frac{1}{2\kappa}\int d\varphi\Bigg[ \int_{-1}^1\mathcal{P}_{u\varphi}(r\to\infty,u)du - \int_{r_h}^{\infty}\mathcal{P}_{r\varphi}(r,u)\Big|_{u=-1}^{u=1}dr\Bigg] \nn\\
&=&\frac{w_2}{4\kappa}A_\varphi|_{u=-1}^{u=1} = \frac{w_2}{2\kappa}\sqrt{\frac{3 \ell +2}{(\ell +1) (\ell +2)}}\Big[p \cos X(r_h) +q \sin X(r_h)\Big]\,.\label{magnetic charge BB}
\eea

\subsection{Unnormal thermodynamic  quantities}

Although the calculations are involved, the evaluation of the electric and magnetic potentials and charges follows standard procedures. In contrast, the computation of the mass and NUT charge remains debated. In this work, we employ the method of \cite{Liu:2022wku} to provide a set of Taub-NUT black hole thermodynamic quantities that satisfy the first law.

\begin{figure}[htbp]
  \centering
  \includegraphics[width=0.6\textwidth]{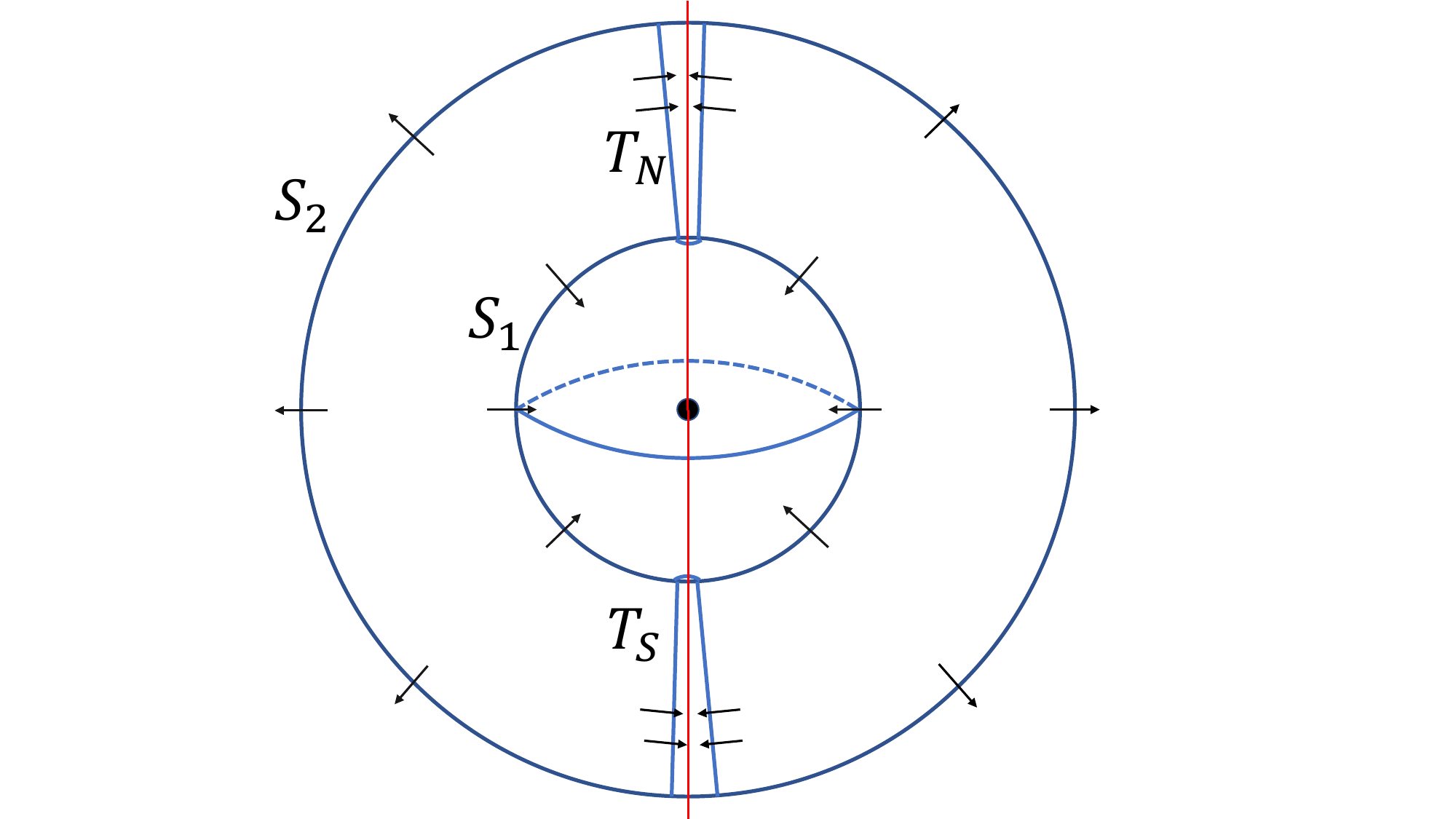}\ \
 \caption{ Owing to the NUT charge, the Wald formalism has singularities at $u = \pm 1$. We select the integration path illustrated below to bypass these singularities. 
}\label{Misner string}
\end{figure}

To simplify the analysis, we introduce a scalar $\Psi$, following \eqref{scalar Psi}, 
\bea 
\Psi&=& -\frac{\sqrt{\ell +1}}{ N (\ell +2)}\Big[
q -q \cos X+p \sin X
\Big] -c_m \,,
\eea 
to eliminate the Dirac string singularity associated with the magnetic charge. Owing to the presence of the Misner string, the Wald formalism diverges at \(u = \pm 1\) during integration. We adopt the integration path illustrated in Fig.~\ref{Misner string}. Equation \eqref{Wald formalism H} can be separated into four distinct parts
\bea 
\delta \mathcal{H}=0=\delta \mathcal{H}_{S_2}+\delta \mathcal{H}_{S_1}+\delta \mathcal{H}_{T_N}+\delta \mathcal{H}_{T_S}\,.\label{Wald formalism taub nut}
\eea 

As discussed in Sec.~\ref{Thermodynamics}, the Wald entropy formula fails to yield a black hole entropy consistent with the first law of thermodynamics $r=r_h$. Consequently, we set \(S_2\) at the black hole horizon and, employing the Wald formalism \cite{Iyer:1994ys}, extract a self-consistent expression for the entropy from \(\delta \mathcal{H}_{S_1}\). We choose the gauge parameters 
\bea 
c_e=-\Phi_e\,,\qquad c_m= -\Phi_m  \,,
\eea 
such that \(\Phi_e \delta Q_e + \Phi_m \delta Q_m\) appears at infinity, in which case \(\delta \mathcal{H}_{r_h}\) yields the black hole entropy
\bea 
\delta \mathcal{H}_{r_h}=T\delta S\,,\qquad \Rightarrow\qquad S=\frac{2 \pi  w_2 (\ell +1)}{\kappa }(r_h^2+N^2)\,.
\eea 

For the Taub-NUT solution, besides the event horizon at \(r = r_h\), there are also two Killing horizons located at the north and south poles \(u = \pm 1\). The two degenerate Killing vectors are
\bea 
u=\pm1:\qquad l_\pm=\partial_\varphi\mp4\Phi_N\partial_t \,.
\eea 
At \(u = \pm 1\), \(l_\pm\) becomes a null vector \(l_\pm^2=0\), which defines the NUT potential \(\Phi_N\)
\bea 
\Phi_N=\frac{N}{2}\,.
\eea 
For $k=0,-1$ case, there is no north or south pole, which means that there is no straightforward way to define the NUT potential by the degenerate Killing vector.

The NUT parameter can give rise to corresponding electric and magnetic charges. Since the integrals associated with the NUT parameter correspond to the paths \( T_{N,S} \) in Fig.~\ref{Misner string}, where \( u = \pm 1 \), the \( dt \wedge dr \) components in \eqref{nut star cal F} and \eqref{nut F} contribute to the two NUT-related electric and magnetic charges
\bea 
Q_{eN}&=&\frac{w_2}{\kappa}\int_{r_h}^{\infty}\mathcal{Q}_{tr}dr=\frac{w_2}{\kappa} \mathcal{A}_t|_{r\to\infty}^{r=r_h}\cr
&=&\frac{w_2\sqrt{\ell +1}}{\kappa  N (\ell +2)}\Big[q-q \cos X(r_h) +p \sin X(r_h)\Big]\,,\cr 
Q_{mN}&=&\frac{w_2}{\kappa}\int_{r_h}^{\infty}\mathcal{P}_{tr}dr=\frac{w_2}{\kappa} A_t|_{r\to\infty}^{r=r_h}\cr
&=&\frac{w_2}{2\kappa N}\sqrt{\frac{3 \ell +2}{(\ell +1) (\ell +2)}}\Big[q \sin X(r_h)+p \cos X(r_h) -p\Big] \,.\label{nut electric magnrtic charge}
\eea 
These charges are also associated with corresponding conjugate thermodynamic potentials
\bea 
\Phi_{eN}&=&\frac{1}{8}l^\mu_-\Big(
A_\mu(u=1)+A_\mu(u=-1)
\Big)\bigg|^{r=r_h}_{r\to\infty}\cr 
&=&\frac{1}{4}\sqrt{\frac{3 \ell +2}{(\ell +1) (\ell +2)}}\Big[q \sin X(r_h)+p \cos X(r_h) -p\Big]\,,\cr
\Phi_{mN}&=&-\frac{1}{8}l^\mu_-\Big(
\mathcal{A}_\mu(u=1)+\mathcal{A}_\mu(u=-1)
\Big)\bigg|^{r=r_h}_{r\to\infty}\cr 
&=&-\frac{\sqrt{\ell +1}}{2  (\ell +2)} \Big[q-q \cos X(r_h) +p \sin X(r_h)\Big]\,.\label{nut electric magnrtic potential}
\eea 
This suggests a relationship between the charges and potentials $Q_{eN}\Phi_{eN}+Q_{mN}\Phi_{mN}=0$ induced by the NUT parameter. We find that there is a relationship between the electric and magnetic charges (\ref{electric charge BB}, \ref{magnetic charge BB}) and the charges induced by the NUT parameter
\bea 
Q_e&=&\frac{w_2 \sqrt{\ell +1}}{\kappa  (\ell +2)}q - 2\Phi_NQ_{eN}\,,\\
 Q_m&=&\frac{w_2}{2\kappa}\sqrt{\frac{3 \ell +2}{(\ell +1) (\ell +2)}}p + 2\Phi_NQ_{mN}\,.
\eea

If we take \(S_2\) to be at infinity, then according to \cite{Iyer:1994ys}, we can set \(c_e = 0 = c_m\) so that the electric and magnetic potentials appear on the horizon, in which case \(-\delta\mathcal{H}_{S_2}\) gives the black hole mass
\bea 
-\delta\mathcal{H}_{\infty}=\frac{  w_2 \sqrt{\ell +1}}{\kappa }\delta m\,.
\eea 
However, as noted in \cite{Liu:2022wku}, if the above expression were taken as the mass, the black hole mass could be either positive or negative. Since a non-positive-definite mass has no physical meaning, the expression represents only a partial contribution to the total mass. This implies that the integrals over \( T_{N,S} \) contribute partially to the total mass
\bea 
&&\delta\mathcal{H}_{T_N}+\delta \mathcal{H}_{T_S}
=-\int_{r_h}^\infty H(r,u)\big|_{u=-1}^{u=1}dr\,,\nn\\
&&H=\frac{\delta p u \sqrt{\ell +1}}{\kappa  (\ell +2) (r^2+n^2)^2}\Big[(r^2+n^2) (p \cos X-q \sin X)+p (n^2-r^2)\Big]\nn\\
&&+\frac{\delta q u \sqrt{\ell +1}}{\kappa  (\ell +2)  (r^2+n^2)^2}\Big[(r^2+n^2) (q \cos X+p \sin X)+q (n^2-r^2)\Big]
-\frac{2  n^2 r u \sqrt{\ell +1}}{\kappa  (r^2+n^2)^2}\delta m\nn\\
&&-\frac{\delta n u \sqrt{\ell +1} (p^2+q^2)}{\kappa  n (\ell +2) \sqrt{3 \ell +2}  (r^2+n^2)^2}\Bigg[
\sqrt{3 \ell +2} (n^2-r^2) \cos X+2 n r \sqrt{(\ell +1)(\ell +2)} \sin X
\Bigg]\nn\\
&&-\frac{\delta n u \sqrt{\ell +1}}{\kappa  n (\ell +2) (r^2+n^2)^3}\Bigg\{
r^4 (p^2+q^2)-3 n^2 r^2 \Big[2 r (\ell +2) (r-m)+p^2+q^2\Big]\nn\\
&&+2 n^4 (\ell +2) \Big[r (6 r-m)+n^2\Big]
\Bigg\}\,.
\eea 
We find that the charges \eqref{nut electric magnrtic charge} and the potentials \eqref{nut electric magnrtic potential} induced by the NUT parameter can be consistently incorporated into the Wald formalism
\bea
-\delta\mathcal{H}_{\infty}
-\delta\mathcal{H}_{T_N}-\delta \mathcal{H}_{T_S}
=\delta M-\Phi_N\delta Q_N-\Phi_{eN}\delta Q_{eN}-\Phi_{mN}\delta Q_{mN}\,.
\eea 
This procedure yields well-defined expressions for the mass \(M\) and the NUT charge \(Q_N\)
\bea 
M&=&\frac{  w_2 \sqrt{\ell +1}}{\kappa }m+2\Phi_NQ_N\,,\cr 
Q_N&=&\frac{2 w_2 N \sqrt{\ell +1}}{\kappa  r_h}\Bigg\{1-\frac{p^2+q^2}{4 N^3 (\ell +2)}\Big[
2 N -r_h\sqrt{\frac{3 \ell +2}{(\ell +1) (\ell +2)}}\sin X(r_h) 
\Big]\Bigg\}\,.
\eea 
The Wald formalism \eqref{Wald formalism taub nut} provides the first law of thermodynamics for the Taub–NUT solution
\bea 
\delta M=T\delta S+\Phi_{e}\delta Q_{e}+\Phi_{m}\delta Q_{m}+\Phi_N\delta Q_N+\Phi_{eN}\delta Q_{eN}+\Phi_{mN}\delta Q_{mN}\,.
\eea 
The Smarr relation is
\bea 
M=2TS+\Phi_eQ_e+\Phi_mQ_m\,.
\eea 

At this stage, we have completed the construction of the dyonic Taub-NUT-like black hole solution and the derivation of its first law of thermodynamics. On one hand, the exact dyonic Taub-NUT-like solution further demonstrates the internal consistency and robustness of the specific EbM model. On the other hand, although the presence of the Taub-NUT charge introduces nontrivial effects on the black hole thermodynamics, the quantities computed via the Wald formalism still guarantee that the first law of black hole thermodynamics holds.

Before concluding this section, we find it appropriate to clarify an important point. To date, a fully satisfactory and universally accepted understanding of the thermodynamics of Lorentzian Taub-NUT spacetimes has not yet been established. In Ref.~\cite{BallonBordo:2020mcs}, the authors employed the quasilocal action method and then computed all thermodynamic quantities using standard statistical mechanics techniques.  They also recognized that the introduction of the NUT parameter gives rise to new charges in the thermodynamics, which \cite{BallonBordo:2020mcs} referred to as horizon electric and magnetic charges. These are precisely the \(Q_{eN}\) and \(Q_{mN}\) calculated in \eqref{nut electric magnrtic charge}. Nevertheless, within that framework, it remains somewhat unclear whether these charges themselves, or instead their conjugate potentials, should be regarded as the appropriate independent thermodynamic variables, and a definitive resolution was not reached.

The literatures \cite{Frodden:2021ces,Godazgar:2022jxm} also computed the thermodynamics using the Wald formalism. However, \cite{Frodden:2021ces} suggested that the mass should be proportional to the parameter \(m\). Yet, the Taub–NUT solution exhibits a peculiar feature: it describes a black object with an event horizon for all real values of \(m\). In other words, for nonzero \(n\), a horizon exists for every \(m \in (-\infty, \infty)\). In \cite{Godazgar:2022jxm}, the NUT potential is defined as being inversely proportional to the NUT parameter \(N\), i.e., $\psi_N=1/(8\pi N)$, which implies that this thermodynamic framework lacks a well-defined \(N \to 0\) limit. However, from the viewpoint of the metric, the Taub–NUT solution smoothly reduces to the Schwarzschild black hole as \(N \to 0\). Hence, this suggests that further clarification of this thermodynamic interpretation may be desirable.

Similar discussions can be found in \cite{Hennigar:2019ive,Wu:2019pzr,Abbasvandi:2021nyv,Wu:2022rmx,Wu:2022mlz,Wu:2022xpp,Wu:2023woq}, where complete thermodynamic frameworks satisfying the first law are presented. The approach used in this work follows essentially the method proposed in \cite{Liu:2022wku}. Despite the variety of perspectives, one fact is clear: the introduction of the NUT parameter adds a new pair of thermodynamic conjugates to the black hole. If the black hole carries electric and magnetic charges, the NUT parameter also introduces the corresponding pair \(Q_{e,mN}\) and \(\Phi_{e,mN}\). Although the number of thermodynamic variables increases, the integration constants remain unchanged, implying that \(Q_{e,mN}\) and \(\Phi_{e,mN}\) are not independent. Moreover, for a rotating black hole, differences between the north and south poles emerge, and the thermodynamic quantities associated with the NUT parameter split accordingly.

Overall, these considerations indicate that the thermodynamics of Lorentzian Taub-NUT spacetimes remains a subtle and actively investigated subject. The successful application of our approach to the dyonic Taub-NUT black hole in Einstein-bumblebee gravity, viewed as a representative vector-tensor gravity theory, provides a concrete and consistent example within this broader context, and may serve as a useful reference for future studies of NUT-charged solutions in modified gravity theories.

\section{ Generalization to higher dimensions} \label{highdim}

In previous sections, we obtained the dyonic RN-like black hole solutions in the four-dimensional EbM theory, and derived the first law of thermodynamics for these black holes. Now in this section, we will generalize these results to arbitrary even dimensions $D= 2+ 2n$.

\subsection{High dimensional dyonic RN-like solutions}

The general ansatz for dyonic topological black holes in $D = 2 + 2n$ dimensions is given by
\bea
ds^2 &=& - h(r) dt^2 + f(r)^{-1} dr^2 + r^2 \sum_{i=1}^n d\Omega_{i, k}^2  \,, \label{metric3} \\
 F &=& -\phi '(r) dt \wedge d r + p \sum_{i=1}^n  dx_i \wedge dy_i \,, \\
 B &=& b_r(r) dr \,, 
\eea
with
\be
d\Omega_{i, k}^2 = \frac{dx_i^2}{1-k x_i^2} + (1-k x_i^2) dy_i^2 \,.
\ee
The solutions for are
\bea
h &=& k - \frac{m}{r^{2 n-1}} + \frac{(2 n-1) q^2 }{2 \left(2 n + \ell \right) r^{2 (2 n-1)} } + \frac{n (2 n-1) \left(1+\ell \right) p^2  }{2(3-2 n) \left(2 n+ (2 n + 1)\ell \right) r^2} \,, \nn \\
f &=& \frac{h}{(2 n-1)(1 + \ell)} \,, \quad \phi = \frac{q}{r^{2 n-1}} \,, \quad b_r = \frac{b}{\sqrt{f}} \,, \label{hdsol}
\eea
where 
\be
\gamma_1 = \frac{\gamma}{4 (2 n + (2 n + 1)\ell)} \,,\quad  \gamma_2 = -\frac{2 n^2 ( 1+\ell ) \gamma }{(2 n + \ell) (2 n +(2 n + 1)\ell)} \,.
\ee
The horizon topology now becomes ${\cal M}_2 \times{\cal M}_2 \times \dots \times {\cal M}_2  $, where ${\cal M}_2$ can be sphere, torus, or hyperbolic 2-space.

\subsection{Thermodynamics}

Here we present the thermodynamical properties of the $D = 2 + 2n$-dimensional dyonic RN-like topological black holes
\bea
&&m = k r_h^{2 n -1} - \frac{n (2 n -1) (\ell +1) p^2 r_h^{2n -3} }{2 (2 n -3) ((2n +1) \ell +2 n)}+\frac{ (2 n -1)q^2 }{2 (\ell +2 n) r_h^{2 n -1}} \,. \nn \\
&&T = \frac{h'}{4\pi\sqrt{(2 n -1)(1+\ell)}} = \frac{k \sqrt{2 n-1}}{4 \pi r_h \sqrt{\ell +1}}-\frac{  n \sqrt{2 n-1} p^2 \sqrt{\ell +1}}{8 \pi  r_h^3 (2 n+(2 n+1) \ell )}-\frac{ (2 n-1)^{3/2} q^2 }{8 \pi  \sqrt{\ell +1} (2 n+\ell ) r_h^{4 n-1}} \,,\nn \\
&&Q_e = \frac{n \sqrt{(2 n -1)(\ell +1)} q w_2^n}{(\ell + 2 n)\kappa} \,, \quad \Phi_e = \frac{q }{r_h^{2n-1}} \,,\nn \\
&&Q_m = \frac{n p w_2}{2 \kappa} \,,\quad \Phi_m = -\frac{2 n \sqrt{2 n-1} p (\ell +1)^{3/2} w_2^{n-1} r_h^{2 n-3}}{(2 n-3) ((2 n+1) \ell +2 n)} \,,\nn \\
&&M = \frac{n m w_2^n}{\sqrt{2 n -1} \kappa} \sqrt{\ell +1} \,,\quad S = \frac{2\pi r_h^{2n}(\ell+1) w_2^n}{\kappa} \,.
\eea
The differential version of first law of black hole thermodynamics can also be written as Eq.~(\ref{1stlaw}), while the integral form reads
\be
M = \frac{2n}{2n-1} T S  +\Phi_e Q_e + \frac{1}{2n-1}  \Phi_m Q_m \,. 
\ee
For the case with $k= 0$, the generalized Smarr relation takes the form
\be
M = \frac{2n}{2n+1} (T S  +\Phi_e Q_e) +\frac{2}{2n+1}\Phi_m Q_m \,.
\ee
All forms of the first law of black hole thermodynamics in higher dimensions remain consistent with those in GR.

\subsection{Some explicit examples}

In the purely electric case, i.e., $p=0$, we can first set $k=0$ and then rewrite $\sum_{i=1}^n d\Omega_{i,k}^2$ as $d\tilde{\Omega}_{D-2,k}^2$,  where $k=1,0,-1$ correspond to $(D-2)$-dimensional spherical, planar, and hyperbolic geometries, respectively. The arbitray $D$-dimensional electrically charged RN-like topological black hole solution is given by
\bea
h &=& k - \frac{m}{r^{D-3}} + \frac{(D-3) q^2 }{2 \left((D-2) + \ell \right) r^{2 (D-3)} }  \,, \nn \\
f &=& \frac{h}{(D-3)(1 + \ell)} \,,  \quad \phi = \frac{q}{r^{D-3}} \,, \quad b_r = \frac{b}{\sqrt{f}} \,,
\eea
where 
\be
\gamma_1 = \frac{\gamma}{4 ((D-2)+ (D-1)\ell)} \,,\quad  \gamma_2 = -\frac{(D-2)^2 ( 1+\ell ) \gamma }{2((D-2) + \ell) ((D-2) +(D-1)\ell)} \,.
\ee
In the purely magnetic case, i.e., $q=0$, the only changes in Eq.~(\ref{hdsol}) are
\be
h = k - \frac{m}{r^{2 n-1}} + \frac{n (2 n-1) \left(1+\ell \right) p^2  }{2(3-2 n) \left(2 n+ (2 n + 1)\ell \right) r^2} \,, \quad \phi = 0 \,. 
\ee

\section{Conclusion} \label{conclusion}

In this paper, we construct four-dimensional dyonic RN-like black holes with general topological horizons within Einstein-bumblebee gravity, one of the simplest vector-tensor theories realizing spontaneous Lorentz symmetry breaking. We then investigate their thermodynamic properties and employ the Wald formalism to compute the conserved mass and entropy, thereby establishing the first law of black hole thermodynamics. Furthermore, we extend the static dyonic solution to the Taub-NUT case, revealing its nontrivial black hole thermodynamics. Finally, we generalize the four-dimensional dyonic RN-like results to higher dimensions. These results provide a concrete framework for further studies of the effects of spontaneous Lorentz symmetry breaking in black hole physics and astrophysics.

Given that Einstein-bumblebee gravity admits GR-like exact black hole solutions involving an increasingly rich set of physical parameters, including mass, electric charge, magnetic charge, and the NUT charge, this highlights the advantages of employing this framework to investigate spontaneous Lorentz symmetry breaking. On the other hand, as emphasized in section~\ref{solutiond4}, although imposing a vanishing potential fixes the functional form of the bumblebee vector $B_\mu$ and reduces the resulting gravitational field equations to second order, the theory remains intrinsically a higher-derivative gravity. As in other higher-derivative gravity theories, the fact that one of the bulk geometries resembles---or even coincides with---its GR counterpart (for instance, many higher-curvature gravities admits the Schwarzschild black hole as an exact solution~\cite{Lu:2015cqa,Li:2017ncu,Liu:2020yqa,Li:2023vbo,Liu:2024wvw,Li:2025gna}) does not imply that the boundary sector must exhibit the same trivial structure. Indeed, the computation of the Noether mass charge via the Wald formalism already demonstrates a clear distinction between Einstein-bumblebee gravity and GR at the boundary level. This motivates extending the present black hole solutions to include a cosmological constant and, through holographic methods, exploring the potential new effects such boundary differences may induce in the dual conformal field theory. Such an analysis could provide new insights into the structure and physical implications of vector-tensor theories of gravity.

Additionally, in a recent work~\cite{Yu:2025odj}, the authors investigated the implications of the Noether mass for observational constraints on spontaneous Lorentz violation, using another Lorentz-violating gravity---Einstein-Kalb-Ramond gravity---as an illustrative example. As shown in that analysis, although different mass definitions lead only to subleading corrections in weak-field observational tests of the Lorentz-violating parameter, the situation would be significantly different in strong-field regimes, such as those involving black holes or neutron stars where relativistic effects are pronounced. In such contexts, constraining the Lorentz-violating parameter using the Noether mass may yield non-negligible deviations compared with previous studies that relied predominantly on the Komar mass or the Arnowitt-Deser-Misner mass.

\begin{acknowledgments}
We are grateful to Yu-Qi Chen, Xing-Hui Feng, Zhi-Chao Li, Hai-Shan Liu, Yuxiao Liu, Yanzhao Lu, H. L\"u and Hongwei Yu for useful discussions. 
S.L., and L.L. were supported in part by the National Natural Science Foundation of China (No. 12105098, No. 12481540179, No. 12075084, No. 11690034, No. 11947216, and No. 12005059) and the Natural Science Foundation of Hunan Province (No. 2022JJ40264), and the innovative research group of Hunan Province under Grant No. 2024JJ1006. L.M.~is also supported in part by National Natural Science Foundation of China (NSFC) grant No.~12447138, Postdoctoral Fellowship Program of CPSF Grant No.~GZC20241211 and the China Postdoctoral Science Foundation under Grant No.~2024M762338.

\end{acknowledgments}


\end{document}